\documentclass{article}

\usepackage{arxiv}
\usepackage{float}
\usepackage[utf8]{inputenc}
\usepackage[T1]{fontenc}
\usepackage{hyperref}
\usepackage{url}
\usepackage{booktabs}
\usepackage{amsfonts}
\usepackage{amssymb}
\usepackage{nicefrac}
\usepackage{microtype}
\usepackage{amsmath}
\usepackage{graphicx}
\usepackage{tabularx}
\usepackage{array}
\usepackage[numbers,sort&compress]{natbib}
\usepackage{doi}

\graphicspath{{figures/}}
\newcolumntype{Y}{>{\raggedright\arraybackslash}X}
\title{Save 2050: A Planetary-Scale Collective Prediction System for the Singularity Crisis}
\date{August 2026}
\author{
    \href{https://orcid.org/0000-0001-7402-7482}{\includegraphics[scale=0.06]{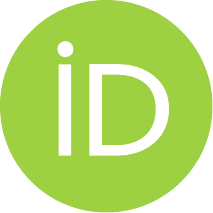}\hspace{1mm}Jiang Zhang}
    \thanks{School of Systems Science, Beijing Normal University}
    \thanks{Swarma Research} \\
    \texttt{zhangjiang@bnu.edu.cn} \\
    \And
    \href{https://orcid.org/0009-0007-9062-0383}{\includegraphics[scale=0.06]{orcid.pdf}\hspace{1mm}Bing Yuan}
    \footnotemark[2] \\
    \texttt{yuanbing@swarma.org} \\
    \And
    \href{https://orcid.org/0009-0006-9084-8356}{\includegraphics[scale=0.06]{orcid.pdf}\hspace{1mm}Qian Zhang}
    \footnotemark[2] \\
    \texttt{zhangqian@swarma.org}
}

\renewcommand{\shorttitle}{Save 2050}
\hypersetup{
  pdftitle={Save 2050: A Planetary-Scale Collective Prediction System for the Singularity Crisis},
  pdfauthor={Jiang Zhang, Bing Yuan, Qian Zhang},
  pdfsubject={Planetary-scale collective prediction and the singularity crisis},
  pdfkeywords={collective prediction, prediction aggregation, long-horizon resolution, collective intelligence}
}

\begin{document}

\begin{center}
  \raggedright
  \includegraphics[height=0.75cm]{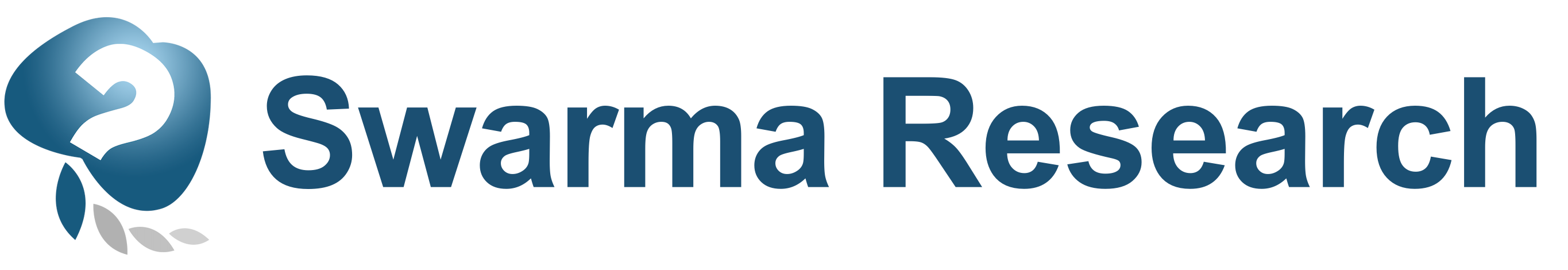}
  \par
\end{center}

\maketitle

\begin{abstract}
The Anthropocene mode of development is driving civilization toward a "singularity crisis": artificial intelligence (AI) is rapidly approaching general intelligence (AGI) with escalating risks of losing control, while unrestrained economic growth generates super-exponential growth of entropy production that pushes the Earth system toward its tipping points. This paper proposes the "Save 2050" initiative: a distributed planetary-scale collective prediction system that aggregates judgments about the future from humans and AI through open registration, crowdsourcing, and incentive mechanisms, and integrates them via large-scale simulation into inspectable "predicted worlds," enabling humanity to systematically \emph{see} the future for the first time. We argue for the initiative's feasibility along four dimensions---the maturation of AI forecasting, human collective intelligence and the institutional environment, supporting progress in related fields, and societal demand. We then identify three key enabling technologies: long-horizon automated resolution, simulation-based prediction aggregation, and reflexivity governance. We analyze potential risks---including reflexivity, cognitive monoculture, narrative capture, and regulatory and ethical concerns---together with mitigation strategies, and we outline a phased roadmap with open problems. The initiative's primary goal is not to intervene in the future, but to make the future visible, discussable, and co-writable.
\end{abstract}

\keywords{collective prediction \and prediction aggregation \and long-horizon automated resolution \and self-fulfilling prophecy \and collective intelligence \and singularity crisis}

\section{Introduction}

Humanity is entering a critical period in which several distinct but coupled trajectories may approach singularity-like limits. Artificial intelligence is progressing toward increasingly general forms of reasoning and agency, reviving the question of an intelligence or capability singularity \citep{good1965,yudkowsky2007,chalmers2010,yudkowsky2013}. At the same time, urban-economic development continues to concentrate population, innovation, wealth, energy use, and entropy production in large cities~\citep{west2017scale,bettencourt2007}, while climate tipping elements and ecological systems exhibit nonlinear and potentially abrupt transitions \citep{lenton2019,armstrongmckay2022,sole2022}. The ``singularity crisis'' considered here is therefore not the claim that these events are identical, but the hypothesis that their coupled trajectories may enter the same historical window.

The urban component of this argument comes from the city-growth singularity discussed by Bettencourt, West, and their collaborators. Many socioeconomic and innovation-related quantities scale superlinearly with city population, approximately $Y(N) \propto N^{\beta}$ with $\beta>1$, whereas infrastructure tends to scale sublinearly \citep{bettencourt2007,bettencourt2010}. In the associated dynamical picture, increasing urban scale accelerates innovation and economic activity, so that city growth can become super-exponential and approach a finite-time singularity. A technological or organizational revolution may postpone that singularity by opening a new growth regime, but the system then enters another round of accelerated expansion; if the intervals between such transformations continue to shrink, postponement does not amount to a final solution \citep{bettencourt2007,west2017scale}. Because large cities are disproportionate contributors to economic and technological output, this urban mechanism can feed into super-exponential GDP growth. Technology is both an input to and an outcome of GDP growth, making it plausible---though not established---that an AI capability singularity and an urban-economic singularity could occur in the same historical window. These remain different phenomena, coupled rather than identical \citep{bettencourt2010,west2017scale}.

The coupling becomes consequential when energy and planetary feedbacks are included. GDP growth under present production systems requires energy and material throughput, with energy conversion accompanied by dissipation, waste heat, and entropy production. The carbon and energy costs of AI are already documented at the levels of model development and deployment: training and running deep-learning systems can carry substantial carbon footprints \citep{dhar2020carbon,strubell2019energy}, while the compute used in the largest training runs has grown rapidly \citep{openai2018aicompute}. Recent assessments also emphasize the growing energy requirements of AI and data-centre infrastructure \citep{iea2025energyai}. Considering the hierarchical pyramid structure of energy consumption~\citep{odum1988selforganization}, the sequential development of cutting-edge technologies will indirectly drive multiplicative energy consumption and entropy production at the base of the pyramid, thereby leading to environmental degradation. In the limiting case, persistent exponential growth in technological energy consumption can threaten the habitability of an Earth-like planet through waste heat \citep{balbi2025}. Together with climate tipping points and ecological complexity, these constraints imply that the AI, urban-economic, and environmental singularities may be synchronized by shared flows of capability, GDP, energy, and matter, without being reducible to one another \citep{lenton2019,armstrongmckay2022,sole2022}.

What response is available to a civilization facing such a coupled crisis? Wong and Bartlett's asymptotic-burnout hypothesis identifies two broad paths. The first is interstellar migration: civilization might continue expanding beyond Earth and use access to a larger energy and material base to escape the local crisis. Yet this route may be very narrow. The energy available to humanity, the engineering difficulty of interstellar expansion, and the possibility that waste heat and other thermodynamic constraints accumulate faster than expansion can proceed all limit the feasibility of treating space colonization as a general solution \citep{wong2022,balbi2025}. The second path is a transition toward \emph{homeostatic awakening}: replacing an indefinitely accelerating material-growth trajectory with a system capable of maintaining the conditions for its own long-term persistence. This civilization-scale extrapolation remains contestable, however, and related work emphasizes that it may be altered by research, foresight, diversity, and future-oriented governance \citep{wong2022,jackson2024wong}.

Homeostatic awakening is not a new idea. The Gaia tradition views the Earth as a coupled system in which biological, atmospheric, geological, and climatic processes can generate stabilizing feedbacks; recent work develops this intuition through dynamic planetary processes and Gaia signatures, ecological resilience, and active inference across living systems \citep{wong2024earth,sole2022,montgomery2023gaia}. The global-brain tradition approaches the same problem from the informational side: human beings, institutions, communication networks, and machines can be organized as a distributed cognitive system that senses, learns, and coordinates at planetary scale \citep{heylighlen2011}. Taken together, these perspectives suggest that homeostatic awakening requires more than a change in individual values. It requires a global self-regulating and self-managing system---in a stronger computational sense, a planetary self-referential system that can construct a model of the Earth and its societies, use that model to anticipate possible futures, and feed the resulting knowledge back into collective regulation. Save 2050 is proposed as a key step toward this capacity: the construction of a planetary-scale system for predicting the future together.

Among the most relevant engineering precedents are Earth-system digital-twin initiatives---most notably the European Union's Destination Earth project~\citep{bauer2021,voosen2020destinationearth}---and Fei-Yue Wang's theory of parallel intelligence~\citep{wang2016parallel,wang2018parallel}. Bauer, Stevens, and Hazeleger describe a digital twin of Earth as an interactive combination of high-resolution Earth-system models, continuous observations, and data assimilation, rather than a static visualization or a large data archive \citep{bauer2021}. Wang's parallel-intelligence framework complements this physical-model perspective by combining artificial systems, computational experiments, and parallel execution, so that multiple ``parallel worlds'' can be generated, compared, and used to guide action \citep{wang2016parallel,wang2018parallel}. Its extension to social digital twins explicitly addresses virtual--real interaction and the representation of complex social systems \citep{wang2020parallelsocieties}. Save 2050 draws on both traditions but changes the center of gravity. Its primary technical problem is not only to simulate the Earth more accurately; it is to aggregate the distributed predictions of humans, institutions, AI models, and domain simulators into a continuously queryable and correctable representation of possible futures.

This distinction explains why Save 2050 is not simply another digital-twin project. A digital twin of Earth is primarily organized around a physical object, its sensor data, and the calibration of a dynamical model. Save 2050 is organized around a distributed prediction network: its basic data are claims about future states and events, its central operation is large-scale prediction aggregation, and its public interface is a search-engine-like query over possible futures. Simulation remains essential, but mainly as the computational environment in which heterogeneous predictions are coupled, scenarios are generated, and consequences are made inspectable. The proposed platform therefore aims to become an Internet-scale memory and model of humanity's expectations about the future, while preserving the plurality of scenarios and the ability to revise or fork the underlying representations.

\begin{figure}[H]
\centering
\includegraphics[width=\linewidth]{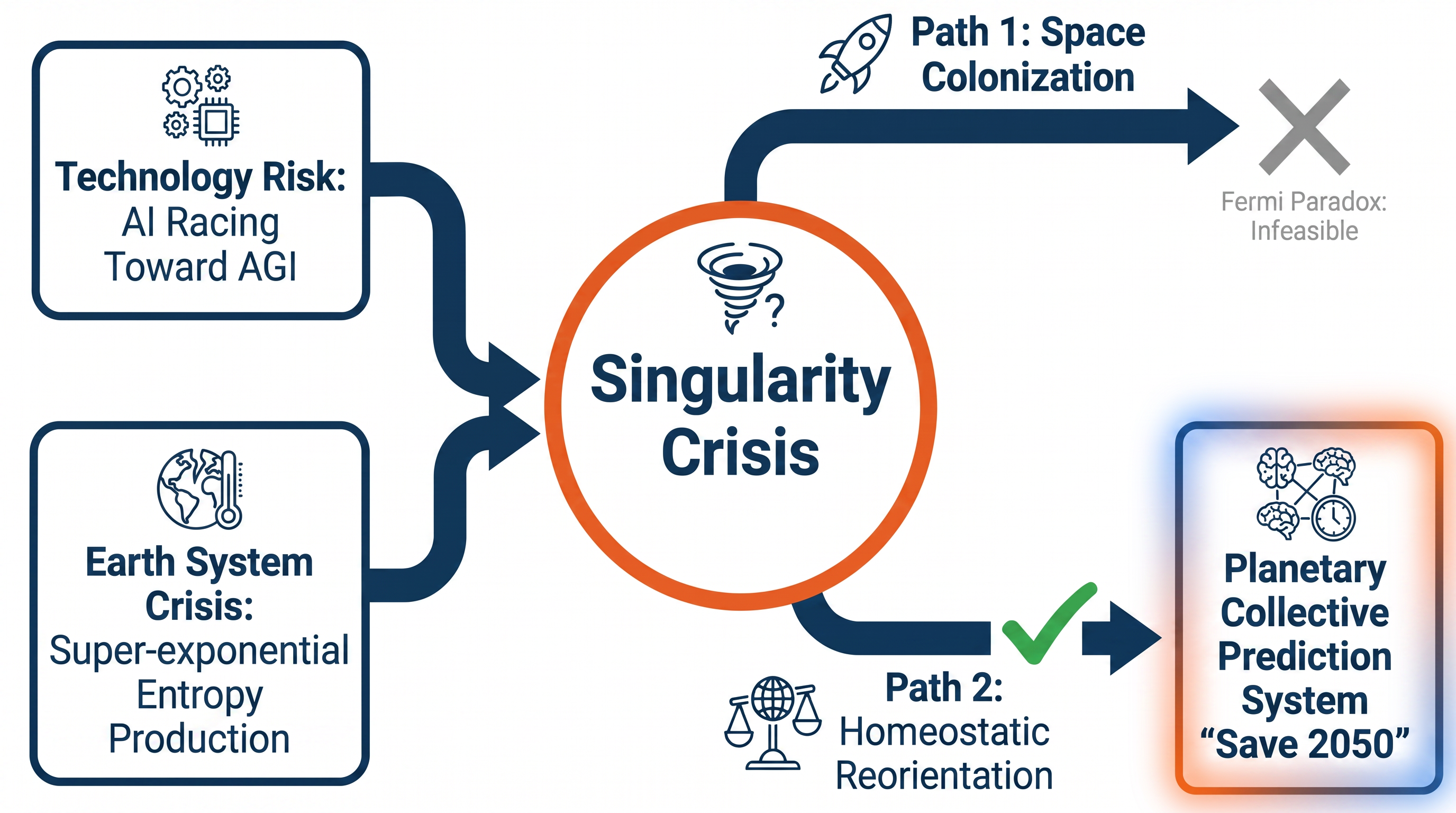}
\caption{The dual structure of the singularity crisis and the logical position of the ``Save 2050'' initiative.}
\label{fig:save2050-1}
\end{figure}

Against this background, Save 2050 proposes a distributed planetary-scale collective prediction system, as shown in Figure~\ref{fig:save2050-1}. Its purpose is not to assert a single inevitable future or to intervene blindly in a system it does not understand. It is to make coupled trajectories visible, discussable, and revisable by aggregating judgments from humans and AI, resolving predictions over long horizons, and embedding them in inspectable simulations. Current AI forecasting and generative-agent systems make parts of this program technically plausible \citep{bubeck2023,metaculus2026,park2023}; digital-twin and ensemble-forecasting research provides a related methodological foundation \citep{bauer2021,kalnay2003}. The initiative is intended as an institutional and epistemic response to a singularity window: before the system is forced into an irreversible transition, society needs a shared instrument for seeing how its own choices shape the future.

The remainder of this paper is organized as follows. Section 2 introduces the distributed Save 2050 platform through four linked functions: prediction aggregation, prediction query, feedback, and prediction restart. Section 3 argues for the initiative's feasibility from technological and societal perspectives. Section 4 analyzes three key enabling technologies. Section 5 examines potential risks and mitigation strategies. Section 6 presents a discussion, a phased roadmap, and open problems.

\section{The Save 2050 Platform: A Distributed Planetary-Scale Prediction System}

\begin{figure}[H]
\centering
\includegraphics[width=\linewidth]{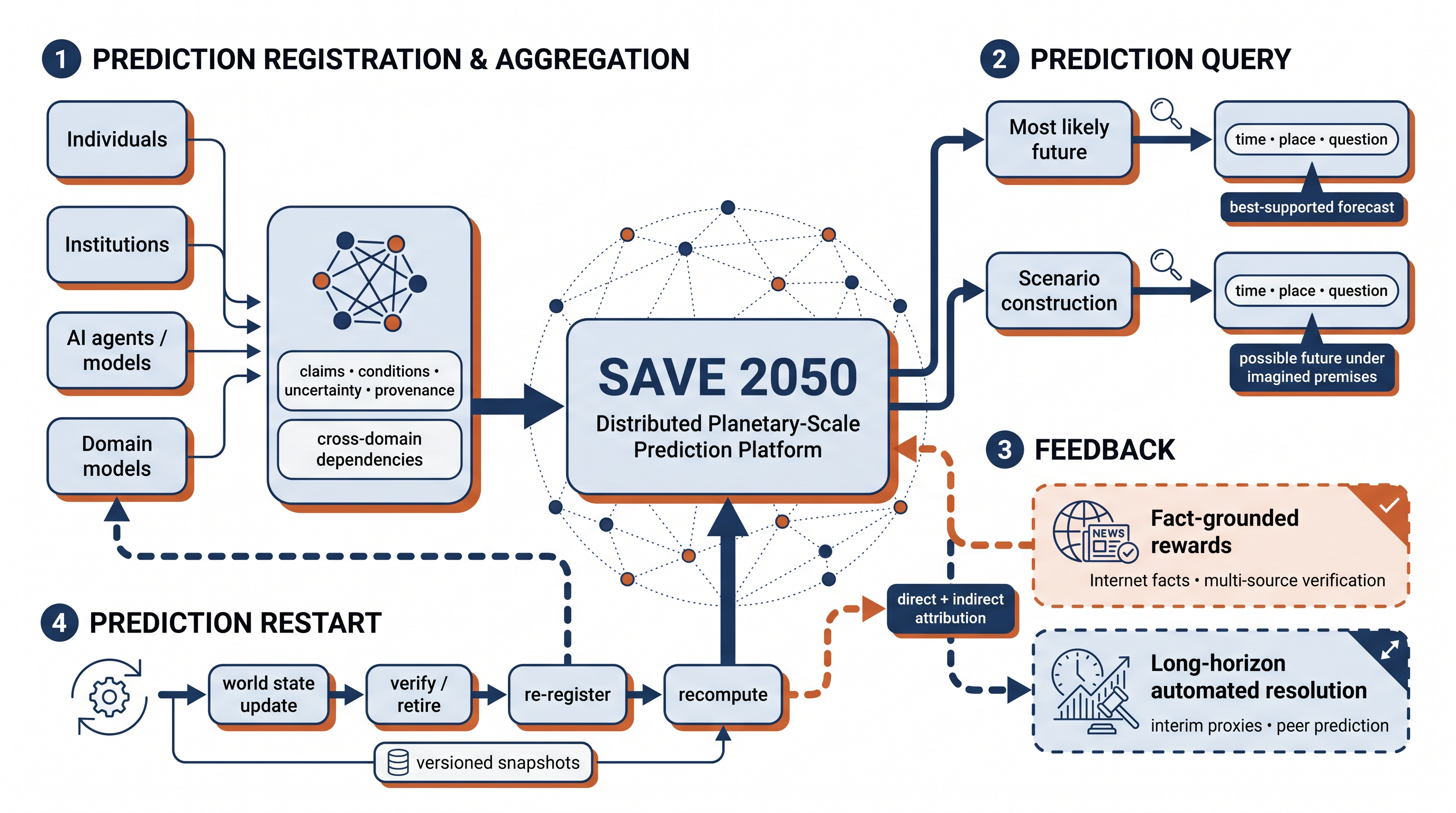}
\caption{Composition and recursive information flow of the distributed Save 2050 platform. Solid arrows indicate forward data and prediction flow; dashed arrows indicate feedback and restart/update flows.}
\label{fig:save2050-2}
\end{figure}

Save 2050 operationalizes the homeostatic and self-referential response outlined in the Introduction as a distributed computing platform for prediction aggregation, prediction query, feedback, and prediction restart. The four functions and their recursive relationships are summarized in Figure~\ref{fig:save2050-2}. Its distributed character is essential: forecasts, evidence, computation, and evaluation need not be held by one server, institution, model, or geographical location. Instead, heterogeneous participants and computational nodes can contribute forecasts, maintain specialized models, replicate evidence, and independently verify results while interoperating through common protocols. The platform is designed as a closed but revisable loop. It does not assume that one model or one institution possesses a complete view of the future; it treats the future as a structured space of competing possibilities that can be registered, propagated, queried, tested, and updated.

\subsection{Prediction Aggregation}

Prediction aggregation begins with an open registration mechanism. Any individual, institution, research group, company, public agency, or AI agent or model may register a prediction after making it. A registration records the claim in a machine-readable form, together with its time horizon, spatial scope, variables, probability or uncertainty range, conditions, provenance, author or model identity, and the resources or evidence on which it depends. Registration is not merely a way to archive forecasts. It makes a forecast an active node in a distributed prediction network: later predictions may cite it, condition on it, refine it, contradict it, or use it as an input. The system should retain these relationships explicitly rather than treating all forecasts as independent votes.

Predictions from different disciplines and systems can therefore become interwoven. A forecast about energy policy may influence forecasts about emissions, prices, migration, climate risk, and political responses; those forecasts may in turn alter the assumptions of the original forecast. The aggregation layer represents these cross-domain dependencies as a directed and versioned network, while distinguishing statistical correlation, informational influence, and causal assumptions. The detailed implications of this reflexive structure are discussed in Section 4.2. Distributed registration also preserves provenance and plurality: no single aggregator needs to erase disagreement in order to produce a system-level forecast.

The system then performs three related operations. First, AI tools normalize heterogeneous submissions into propositions with explicit dates, locations, metrics, uncertainty ranges, and conditional assumptions. Second, a distributed set of aggregation and simulation services couples registered predictions so that cross-domain consequences become visible. Third, the platform produces multiple coherent trajectories rather than forcing every input into one point estimate. Ensemble forecasting and data assimilation provide an established physical analogue: multiple trajectories represent a distribution over possible states, observations contract that distribution, and the updated ensemble evolves forward again \citep{kalnay2003,bauer2021}. Save 2050 extends this pattern to human judgments and AI-generated forecasts, with the registered prediction network and its aggregation---not simulation alone---as the primary source of system-wide intelligence.

The objective is a prediction that is as accurate, detailed, and conditionally informative as possible. A single unconditional forecast can hide disagreement and create false certainty; the platform should preserve alternative scenarios, identify the assumptions under which each scenario holds, and expose the contribution and reliability of different sources. Techniques for combining probability predictions, including logit-space aggregation, offer useful building blocks, but the planetary setting also requires explicit treatment of correlation, common data sources, dependence between registered predictions, and model monoculture \citep{satopaa2014}. The result is a continuously updated computational representation of possible futures rather than a single authorized narrative.

\subsection{Prediction Query}

Once the distributed registration, aggregation, and simulation layers are in place, Save 2050 can be used in a manner analogous to today's Internet search engines. A search engine retrieves and organizes information about what has already happened or what is currently known; Save 2050 would retrieve and organize structured expectations about what may happen. A user could specify a future time and place and ask a concrete question such as: ``What will the weather and rush-hour traffic in Beijing be like on the evening of July 5, 2050?'' The platform would return the most likely future calculated from the registered predictions and their aggregation, together with probability ranges, causal assumptions, provenance, and alternative scenarios.

The query interface should support two distinct modes. In the first, the query asks for the platform's current best-supported or most likely future. The answer is a retrieval and synthesis of the distributed prediction field: the system weighs registered forecasts, accounts for their dependencies, incorporates observations and model outputs, and reports both its central projection and the uncertainty or disagreement around it. This mode should not conceal the fact that the result is conditional on the currently registered knowledge and assumptions.

In the second mode, the querier imagines a scenario and asks the system to construct a possible future from it. For example, the querier might specify a rapid transition to a new energy technology, a severe geopolitical disruption, or a particular urban-planning decision. The system can use its registered predictions, historical knowledge, domain models, and generative models to generate a coherent trajectory under those premises. This is a scenario-construction mode, not a claim that the generated future is the most probable one. The interface should label the distinction clearly and show which assumptions were supplied by the querier, which were inherited from registered predictions, and which were generated by the system.

Either mode may return a narrative description, a probability distribution, a map, a time series, or a comparison of policy choices. A user may inspect how the result changes under different emissions pathways, infrastructure investments, migration patterns, or energy systems. Generative-agent simulations show how language models can support structured social environments with memory, planning, and interaction \citep{park2023}; digital-twin research shows how observations and models can be coupled for continuous state estimation \citep{bauer2021}. Save 2050 combines these capabilities with the much larger input space of crowdsourced human and machine predictions.

The query interface is also the platform's principal mechanism for making planetary intelligence public. Rather than leaving forecasts inside expert institutions or proprietary models, it allows different groups to inspect the assumptions behind a predicted world, compare competing trajectories, and ask how the future changes when a particular condition is altered. In this way, prediction becomes a shared object of inquiry and coordination rather than an opaque output delivered by a single authority.

\subsection{Feedback and Long-Horizon Automated Resolution}

The feedback function has two complementary tracks. The first is a fact-grounded reward mechanism. As time passes, some registered predictions become directly verifiable: weather observations, energy production, economic indicators, infrastructure changes, or the occurrence of a specified public event can be compared with the registered claim. Today's Internet can serve as the initial global fact aggregator for this purpose. News reports, official releases, scientific observations, sensor feeds, public records, and other online sources can be collected automatically; independent sources can then be cross-checked to reduce the risk of relying on one report, while provenance and uncertainty remain attached to the verification.

When a prediction is verified, the system can reward its registered source through money, reputation, access privileges, or other forms of recognition. The reward should reflect calibration, specificity, timing, and usefulness under the stated conditions, rather than only binary correctness. The registration graph also permits indirect reward transmission. If prediction $A$ materially influenced prediction $B$, and $B$ is later confirmed by facts, then the system can allocate a suitably discounted share of the reward to both $B$ and $A$. Such attribution must be evidence-based and bounded: it should distinguish documented informational influence from mere similarity, avoid rewarding copied errors, and prevent an unbroken chain from distributing the entire reward indiscriminately. The same graph can provide negative feedback when a registered prediction is contradicted or when it systematically leads downstream forecasts astray.

The second track is guided by long-horizon automated resolution. Many claims, especially those extending to 2050, cannot wait decades for a final verdict, and some concern latent or counterfactual states for which no single direct fact will ever provide a complete ground truth. In these cases, the platform can use formalized claims, interim proxy indicators, peer prediction, comparisons with competing forecasters, and mark-to-market or other staged scoring methods. Prediction-market and peer-prediction mechanisms provide relevant building blocks when objective ground truth is delayed or incomplete \citep{hanson1995,wolfers2004,wolfers2006,prelec2004,miller2005}; smart-contract and oracle architectures provide additional components for settlement-ready claims \citep{uma2024}. This track should be explicitly labeled as provisional and mechanism-guided, rather than presented as equivalent to fact-based verification.

The two tracks can inform one another without being conflated. Internet-derived facts provide hard anchors wherever they are available, while long-horizon automated resolution supplies interim signals for claims that cannot yet be directly checked. Both signals can update source reliability and aggregation weights, but the platform should preserve the distinction between an observed fact, a statistically supported inference, a peer assessment, and a model-generated judgment. This separation is necessary for auditability and for preventing a self-reinforcing narrative from being mistaken for reality.

The feedback loop also makes the system self-referential in a practical sense. The platform predicts the world, its predictions influence how people understand and act in the world, and the resulting observations update the platform's own model. This reflexivity is a risk as well as a source of intelligence, so the platform must retain provenance, alternative scenarios, auditability, and the ability to challenge or fork a representation. The governance requirements are developed further in Section 5.

\subsection{Prediction Restart Mechanism}

A prediction platform cannot be initialized once and then left to run unchanged. As time passes, the state of the world is updated: new observations arrive, events are resolved, institutions change, and some assumptions become obsolete. Some registered predictions are automatically verified, some are falsified or rendered irrelevant, and some remain unresolved but acquire new evidence. Consequently, both the predicted future states and the entire population of registered predictions may change. Save 2050 therefore requires a prediction restart mechanism that periodically reconstructs the active prediction field from the latest state of the world.

A restart begins by collecting and verifying new facts from the Internet and other trusted observation channels, using multiple sources and retaining source provenance. Resolved predictions are closed with their verification record and scoring outcome. Predictions that are contradicted, superseded, or no longer meaningful are retired from the active set, while unresolved predictions are revised when their conditions, variables, or evidence have changed. Participants and AI agents may register new predictions in response to the updated state, and existing predictions may be reissued as new versioned claims rather than silently overwritten.

The aggregation layer then recomputes dependencies, reliabilities, scenarios, and future trajectories from the updated state. Each restart should produce a versioned snapshot so that researchers can compare what the platform believed at different times, trace why a forecast changed, and distinguish a genuine improvement from a change caused by new assumptions or altered participation. Old predictions and their outcomes remain part of the historical record, but only the current active version should influence current queries unless a query explicitly requests a past snapshot.

In this sense, prediction restart is the temporal counterpart of distributed computation. The network continuously alternates between registration, propagation, observation, verification, and reconstruction. The Internet supplies an already existing, global-scale infrastructure for collecting reports about major events; as long as an open Internet or successor network continues to exist, its information can support both fact verification and the next round of forecasting. The result is not a fixed prophecy but a living, versioned, and collectively maintained model of possible futures.

\section{Feasibility Analysis}

\subsection{The Maturation of AI Forecasting}

AI has moved from the era of "big data" to the era of "big models"; the next stage will be "grand prediction." This judgment, first made in 2023, is now being vindicated.

In general-purpose forecasting, LLM-driven systems have crossed the threshold of the general public: as of 2026, AI forecasting systems consistently outperform the public in open, forward-looking tournaments, are roughly on par with active human forecasters, and continue to close in on teams of top professionals; multiple capability curves extrapolate to human top-level performance within one to two years \citep{metaculus2026}. Equally important, the "recipe" for effective forecasting systems is now well understood---frontier reasoning models, iterative agentic search, multi-model ensembling, and post-hoc calibration---while the cost of a single high-quality forecast has fallen to the order of one US dollar. Prediction, for the first time, has become a cheap, mass-producible capability.

In domain-specific forecasting, the vision that "every domain will have its own foundation model" was first confirmed in meteorology: machine-learning weather models now outperform traditional numerical weather prediction in medium-range global forecasting accuracy, at negligible inference cost \citep{lam2023,price2024}. Meanwhile, LLM-driven world simulators and agent-based social simulations have, for the first time, made it possible to "put social systems into a simulator" \citep{park2023}.

Admittedly, the difficulty of most complex real-world problems lies not in any single domain but in the interaction and coupling between systems. Improving prediction for complex systems therefore requires fusing predictions across domains and scales---the essence of "grand prediction." If all predictions are translated into language, large language models already provide a viable interface for such fusion. Practice over the past three years, however, has revealed that the true bottleneck has shifted from "fusion" to "resolution": how to formalize vague natural-language predictions and automatically judge their eventual correctness (see Section 4.1).

\subsection{Human Collective Intelligence and the Institutional Environment}

Prediction need not rely on AI alone; it can equally draw on the collective intelligence of human crowds through crowdsourcing and human computation. The effectiveness of human crowd forecasting has long been demonstrated: in large forecasting tournaments, selected and trained teams of ordinary forecasters outperformed experts and intelligence analysts by striking margins \citep{tetlock2015}; and the effectiveness of prediction markets as information-aggregation mechanisms is supported by decades of theory and evidence \citep{wolfers2004,arrow2008}.

More important still is the maturation of the institutional environment. Since 2024, event-contract markets have gone mainstream: platforms such as Kalshi operate as Designated Contract Markets registered with the U.S. Commodity Futures Trading Commission (CFTC), monthly prediction-market volume has reached the tens of billions of dollars, and regulators have initiated dedicated rulemaking \citep{crs2025,kpmg2026}. For the first time, crowdsourced prediction and its incentives possess a compliant, scalable institutional channel.

On the incentive side, Web3 engineering components are likewise in place: oracle networks and futures-style contract templates already allow settlement-ready prediction contracts on arbitrary metrics \citep{uma2024}; and for long-horizon predictions aimed at 2050, the mark-to-market and position-transferability institutions of futures markets can be borrowed, complemented by peer-prediction mechanisms that incentivize honesty in the absence of objective ground truth \citep{prelec2004}, so as to build periodic interim resolution (see Section 4.1).

\subsection{Supporting Progress in Related Fields}

At the planetary scale, digital-twin programs represented by the EU's Destination Earth are extending the paradigm of ensemble forecasting and data assimilation to the entire physical Earth \citep{bauer2021}. This paradigm is, notably, a mature prototype of the "simulation-based prediction aggregation" advocated here: dozens of trajectories evolve in parallel to form a distribution over possible futures, and observations periodically contract that distribution toward the realized state---the full "superposition--collapse--re-evolution" cycle has operated in meteorology for decades \citep{kalnay2003}.

Forecasting for global population, economic, and other systems can likewise be incorporated into the initiative's scope. As early as the 1970s, the World3 model in \emph{The Limits to Growth} attempted to capture cross-domain interactions among population, industry, resources, and environment through system dynamics \citep{meadows1972}. Although such early models fall far short of today's AI-foundation-model-based prediction, their cross-domain, holistic perspective remains highly relevant.

\subsection{Societal Demand and a Sustainable Model}

Societally, prediction is a perennial human need: from I Ching divination to mathematical models to artificial intelligence, prediction has always been a central problem. Traditional forecasting, however, is fragmented---predictions made by different people and institutions never intersect---so the problem has never broken through its limits of scale and accuracy. Just as successfully predicting a few stocks can yield enormous returns, successful prediction creates enormous value; an aggregation platform for prediction is therefore bound to emerge.

Such a platform bears a deep resemblance to the search engine: search engines aggregate the information humanity already has, whereas a prediction-aggregation platform would aggregate humanity's judgments about the future, forming a vast database of future "facts" that every user can query the way they use a search engine today---a "Future Engine." The order-of-magnitude decline in the cost of AI prediction makes this engine commercially sustainable for the first time: when prediction is cheap enough, it can be embedded, like search, into the daily decisions of every organization and every individual. As predictions grow more accurate and participation expands, the system forms a self-reinforcing flywheel.

\section{Key Technologies}

Save 2050 depends on coordinated progress in forecasting science, artificial intelligence, and mechanism design. As shown in Table~\ref{tab:save2050-1}, three technologies occupy pivotal positions; in order of priority, they are: long-horizon automated resolution, prediction aggregation, and reflexivity governance. The three are not independent: resolution provides the platform's "judge," aggregation its "brain," and reflexivity governance answers the question "whom does this brain represent?"

\begin{table}[H]
\centering
\small
\caption{Overview of the three key technologies.}
\label{tab:save2050-1}
\begin{tabularx}{\linewidth}{>{\raggedright\arraybackslash}p{.17\linewidth} >{\raggedright\arraybackslash}p{.23\linewidth} >{\raggedright\arraybackslash}p{.36\linewidth} >{\raggedright\arraybackslash}p{.20\linewidth}}
\toprule
 Technology & Problem addressed & Core idea & Current foundations                                                                               \\
\midrule
 Long-horizon automated resolution & Long-horizon, natural-language predictions cannot be judged; incentives and feedback collapse & Borrow futures institutions: question formalization + mark-to-market interim scoring + transferable positions, with peer prediction as an "artificial anchor" & Event contracts under futures regulation \citep{crs2025}; oracles and contract templates in place \citep{uma2024}       \\
 Simulation-based prediction aggregation & How to fuse heterogeneous, correlated, cross-domain predictions into a coherent whole & Predictions as "wavefunctions" superposed in a simulation platform, "collapsed" into predicted realities, cyclically corrected by observations & Ensemble forecasting and data assimilation mature for decades \citep{kalnay2003}; digital twins expanding \citep{bauer2021}  \\
 Reflexivity governance & Once the platform succeeds, predictions act back on the future: who guides? & The simulation platform as the "organ of representation" of collective consciousness---decentralized, plural, auditable, forkable & Terra incognita: no planetary self-fulfilling system has ever existed                             \\
\bottomrule
\end{tabularx}
\end{table}

\subsection{Long-Horizon Automated Resolution}

Resolution is the lifeline of a prediction platform. A prediction whose correctness cannot be determined can neither incentivize forecasters nor provide feedback signals for aggregation. Traditional prediction markets and crowdsourcing platforms resolve only short-term, digitizable events \citep{wolfers2004,arrow2008}, whereas Save 2050 faces predictions that are mostly expressed in natural language, long-horizon, and vaguely referenced (e.g., "the global order in 2050")---far beyond the resolution capacity of existing platforms.

Institutionally, futures markets offer a ready and profound template. The idea of importing futures mechanisms into markets for ideas and predictions was systematically articulated by Robin Hanson thirty years ago under the name "idea futures" \citep{hanson1995}. Futures can sustain "promises about the distant future" thanks to three institutions: standardized contracts, margining with daily mark-to-market, and freely transferable positions. Mapped onto a prediction platform, these form the skeleton of long-horizon automated resolution:

First, automated formalization of prediction claims, corresponding to contract standardization. Large language models can convert vague natural-language predictions into adjudicable propositions---with explicit dates, metrics, and resolution sources---the prerequisite of all resolution.

Second, periodic interim scoring, corresponding to margin and mark-to-market. A prediction about 2050 cannot wait until 2050 to be resolved---after temporal discounting, the incentive would approach zero. Just as futures traders do not wait for delivery but settle gains and losses daily at market prices, the platform must score long-horizon predictions periodically (mature methodologies exist for interpreting and validating market prices as probabilities \citep{wolfers2006}): using staged proxy indicators and "the consensus of future forecasters" (peer prediction \citep{prelec2004,miller2005}) as temporary substitutes for final truth, converting an unreachable terminal resolution into a continuous, predictable process of interim resolution.

Third, transferability of prediction positions, corresponding to the free transfer of futures positions. When predictions and their associated rights (tokens, reputation, contracts) are freely tradable, forecasters need not "hold to 2050" but can exit and realize value at any time---institutionally circumventing the discounting deadlock.

A fundamental difference must, however, be faced: at expiry, futures prices converge to spot prices, and arbitrage provides the final "anchor"; a prediction about 2050 has no such anchor in the interim, and prices may be dominated by noise and sentiment---the celebrated "noise trader risk" literature showed that long-lived assets lacking interim anchors cannot be corrected by arbitrage \citep{delong1990}. Market institutions alone are therefore insufficient; peer-prediction-style mechanism design is needed as an "artificial anchor," scoring and incentivizing predictions based on their relationships to other predictions when objective truth is absent (e.g., the Bayesian Truth Serum \citep{prelec2004}). In engineering practice, contract templates such as UMA's Long Short Pair already allow futures-style synthetic assets on arbitrary metrics, resolved by an optimistic oracle \citep{uma2024}; and at the regulatory level, the convergence of prediction markets and futures is already underway in the United States \citep{crs2025}. The combination of market institutions and artificial anchors may be the realistic solution to long-horizon automated resolution.

\subsection{Simulation-Based Prediction Aggregation}

If resolution answers the question of "correctness," aggregation answers the question of "wisdom." The aggregation envisioned here is far more radical than weighted averaging, extremization \citep{satopaa2014}, market aggregation, or ensemble learning: it is aggregation via large-scale simulation.

The skeleton of the idea is as follows. First, a large-scale (possibly planetary) simulation platform is required. Any prediction can be placed into this platform and coupled with other events and other predictions to construct complete future scenarios. In this sense, each prediction is like a "wavefunction": not an assertion about the future, but a distribution over possible states of the future world. As predictions accumulate, they "superpose"---coupling with and constraining one another---weaving an ever more complete picture of the future. The aggregator's task is to "collapse" these superposed possibilities, given the input predictions, into one or more "predicted realities": concrete, coherent, inspectable future worlds. When time advances to the predicted moment, actual observations enter the system as feedback and revise the aggregator's simulation---an act of "measurement" collapsing the wavefunction---after which the system evolves forward again from the corrected state, generating new "wavefunctions" of the future.

\begin{figure}[H]
\centering
\includegraphics[width=\linewidth]{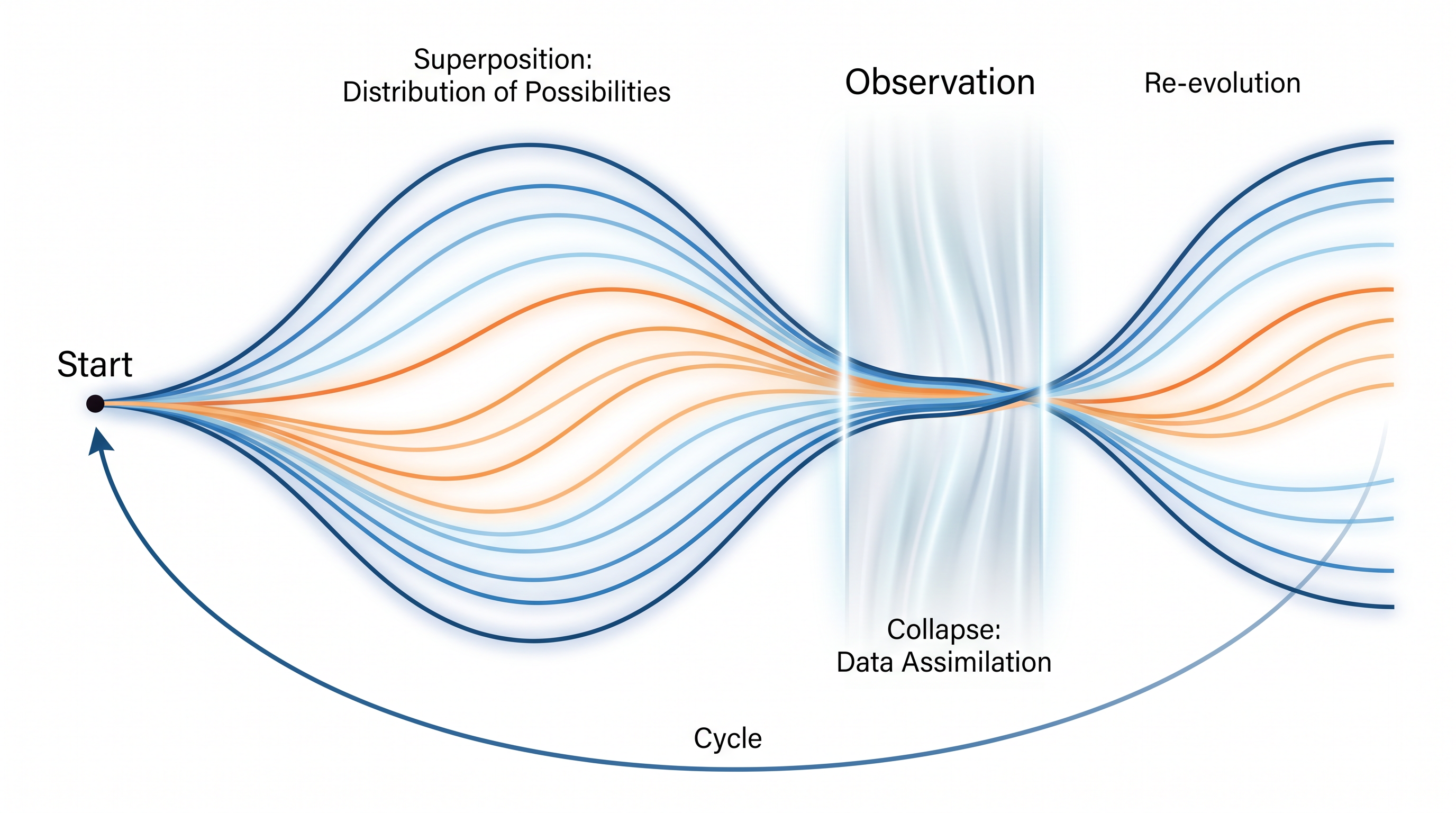}
\caption{The "superposition--collapse--re-evolution" cycle of simulation-based aggregation: parallel trajectories form a distribution of possibilities (superposition); observations contract the distribution toward the realized state (collapse/data assimilation); the ensemble then evolves onward (re-evolution).}
\label{fig:save2050-3}
\end{figure}

This language is not mere metaphor. Ensemble forecasting in numerical weather prediction operates exactly this way: dozens of trajectories with slightly different initial conditions evolve in parallel to form a probability distribution over future atmospheric states; when observations arrive, data assimilation contracts the distribution toward the true state, and the ensemble evolves onward---the complete "superposition--collapse--re-evolution" cycle \citep{kalnay2003} (see Figure~\ref{fig:save2050-3}). In recent years, digital-twin programs have extended this paradigm to the entire physical Earth \citep{bauer2021}; LLM-driven world simulators and agent-based social simulations have made it possible, for the first time, to "put social systems into simulators" \citep{park2023}; and "world simulation" product concepts have begun to appear in industry \citep{metaculus2026}. Simulation-based aggregation thus stands precisely at the convergence of several technological curves.

Three difficulties must nevertheless be overcome. First, the simulation fidelity of social systems: physical systems obey equations, whereas social systems consist of reflexive humans, and simulation errors amplify over time. Second, validation of the simulator itself: a wrong simulator will confidently collapse the future onto wrong realities---an order of magnitude more dangerous than a single wrong prediction---so simulators must be enrolled in the same resolution-and-validation loop as forecasters. Third, the ingestion interface: how is a natural-language prediction translated into a set of perturbations or constraints inside the simulator? This is precisely where large language models serve as the "translation layer."

The long-term vision of this aggregation technology is a "Future Engine": just as search engines aggregated the information humanity already has, a simulation-based aggregation platform will aggregate humanity's imagination and judgment about the future---searchable, interrogable, and evolvable.

\subsection{Reflexivity Governance}

It must be clarified at the outset that the initiative's primary intention is prediction itself---faithfully aggregating humanity's judgments about the future, not intervening in it. Yet the initiative must confront a possibility endogenous to its own success: once planetary-scale prediction becomes real and the public places deep trust in the platform's predictions, prediction ceases to be mere description---it shapes present action, thereby reshaping the future, and may ultimately evolve into a self-fulfilling prophecy. In other words, the more successful the platform, the more unavoidable reflexivity becomes. And reflexivity is a double-edged sword: predictions alter the objects being predicted (Soros's reflexivity \citep{soros1987}); and when a measure becomes a target, it ceases to be a good measure (Goodhart's law \citep{goodhart1975}). When predictions begin to guide humanity toward a "predicted future," we must ask: who is doing the guiding?

The initiative's answer is that the nodes that are truly powerful---and that ought to be powerful---are not the visible, tangible individuals (persons, organizations, machines, large models), but the collective intelligence diffused throughout the global system. Just as a brainwave belongs to no single neuron yet faithfully reflects the state of the brain as a whole, planetary collective consciousness belongs to no person or institution, yet it should be the final arbiter of the future's direction. The world is in crisis today precisely because visible nodes---political institutions, commercial organizations, individual elites---have intercepted and usurped this invisible general will. Reflexivity governance is therefore not about "constraining someone," but about building a mechanism through which collective consciousness can be represented and thereby become dominant.

Governance thus converges with the aggregator described above: the simulation platform that superposes countless predictions and collapses them into "predicted realities" is precisely the organ of representation of collective consciousness---it makes a planetary will that belongs to no one visible, discussable, and accountable for the first time. To govern is to guarantee the fidelity of this organ's representation: the object of governance is not "who said what on the platform," but "whether the platform faithfully renders the emergent patterns of the whole."

It is here, however, that an ancient and dangerous trap must be flagged. Since Rousseau articulated the "general will" (volonté générale) \citep{rousseau1762}, "a will of the whole that belongs to no individual" has been the noblest concept in human politics---and the one most easily usurped: whoever controls the organ that represents the general will may claim to embody it. The organ of representation must therefore never be owned by any single node: it must be a decentralized protocol rather than a company's asset; it must allow multiple competing representers to coexist; it must be auditable, challengeable, and forkable end to end; and it may even explore futarchy-style schemes in which the platform is governed by prediction markets themselves \citep{hanson2013}. Otherwise, "planetary will" becomes the perfect excuse for monopoly.

\section{Risk Analysis}

The greater the initiative's ambition, the graver its failure modes. This section classifies the major risks into five categories and proposes mitigations as shown in Table~\ref{tab:save2050-2}. Most of these risks are not exogenous threats but endogenous to the initiative's success; risk analysis is therefore not a refutation of the initiative but part of its design constraints.

\begin{table}[H]
\centering
\small
\caption{Taxonomy of potential risks and mitigation paths.}
\label{tab:save2050-2}
\begin{tabularx}{\linewidth}{>{\raggedright\arraybackslash}p{.20\linewidth} >{\raggedright\arraybackslash}p{.37\linewidth} >{\raggedright\arraybackslash}p{.37\linewidth}}
\toprule
 Risk & Manifestations & Mitigations                                                                                                                  \\
\midrule
 Reflexivity & Pessimistic self-fulfillment; distortion and trust erosion when predictions become targets (Goodhart effects) & Conditional-probability phrasing; multi-scenario presentation; governance constraining the direction of reflexivity          \\
 Cognitive monoculture & Correlated errors of homologous AI forecasters amplified by aggregation; simulator bias institutionalized & Reward low-correlation forecasters; competing simulators; simulators enrolled in resolution loops                            \\
 Manipulation and capture & Capital distorting consensus; rhetorical manipulation of resolution; operators usurping the "planetary will" & Protocolization and decentralization; manipulation detection and audits; human appeal; community right to fork               \\
 Institutional and regulatory & Gambling classification; cross-border legal conflicts; drift toward betting; compliance risks of long-horizon contracts & Compliance-first design; non-monetary incentives; regulatory sandboxes                                                       \\
 Societal and ethical & Skewed participation causing "representational distortion"; hegemony of future narratives; privacy and data rights & Plural values and multi-scenario representation; low-barrier participation and proof of personhood; data-minimization norms  \\
\bottomrule
\end{tabularx}
\end{table}

\subsection{Reflexivity Risks: Self-Fulfillment and Self-Defeat}

The more successful the platform, the stronger the back-action of predictions, producing two failure modes. The first is pessimistic self-fulfillment: if the "predicted world" presents an unoptimistic future and is widely trusted, it may trigger panic, short-sighted resource grabs, or collective resignation---"if failure is fated, effort is pointless"---thereby realizing the pessimistic future. The second is self-defeat and distortion: when predictions are adopted as targets or policy inputs, the objects being predicted change their behavior accordingly, falsifying the original predictions (Goodhart's law \citep{goodhart1975}); and once predictions repeatedly "fail," the platform's trust capital drains quickly. Mitigations include phrasing predictions as conditional probabilities ("if no action is taken, then \ldots{}") rather than fatalistic assertions; presenting multiple scenarios side by side instead of a single "predicted reality"; and constraining the direction of reflexive effects through the governance mechanisms of Section 4.3.

\subsection{Cognitive Monoculture and Systematic Bias}

When forecasters on the platform rely heavily on homologous AI models and homologous data, their errors are correlated, and the independence assumption underlying classical aggregation fails---aggregation then amplifies shared blind spots rather than canceling them. Worse, if "predicted realities" are generated by a small number of simulators, the simulators' structural biases become institutionalized and are output under the guise of "objective futures." Mitigations include explicitly rewarding low-correlation forecasters in the incentive scheme; requiring multiple independent, competing simulators; and enrolling simulators themselves in the same resolution-and-validation loop as forecasters (Section 4.2).

\subsection{Manipulation and Narrative Capture}

A widely trusted prediction platform is itself an incentive to manipulate: large capital can distort market prices and "consensus"; rhetorical strategies can manipulate natural-language adjudication; and the operator of the "organ of representation" may claim to speak for the planetary will---the trap of "usurping the general will" discussed in Section 4.3. Mitigations include protocolization and decentralization to avoid single-entity control; manipulation detection and anomalous-trading audits; human appeal and review channels in resolution; and preserving the community's "right to fork"---if representation ceases to be faithful, participants can leave with the data and the rules.

\subsection{Institutional and Regulatory Risks}

Prediction markets still face classification as "gambling" in most jurisdictions, and cross-border participation raises complex legal conflicts; even in legalized markets, there are signs of drift toward sports betting---substituting addictive trading for information aggregation \citep{metaculus2026,kpmg2026}. Long-horizon contracts aimed at 2050, if poorly designed, may also cross the red lines of illegal fundraising and securities regulation. Mitigations include compliance-first design within existing regulatory frameworks (such as the futures-regulation channel for event contracts \citep{crs2025}); exploring non-monetary incentives (reputation, honors) to partially substitute for cash settlement; and building "sandbox" experimentation spaces together with regulators.

\subsection{Societal and Ethical Risks}

The deepest risks concern the representativeness of "collective consciousness." If the platform's participant structure is systematically skewed (by geography, language, class, or the digital divide), the so-called "planetary will" is merely the will of privileged groups dressed in the clothing of the whole; and different cultures imagine a "good 2050" differently, so a single "predicted world" could become a hegemony of future narratives. The massive data required by planetary prediction also implicates privacy and data rights. Mitigations include replacing single aggregate metrics with plural values and multi-scenario representations; ensuring broad and genuine participation through low-barrier onboarding and Sybil-resistant proof of personhood; and establishing data-minimization and purpose-limitation norms at the data layer.

\section{Discussion and Outlook}

\subsection{Relation to Existing Paradigms}

Save 2050 can be seen as a distributed convergence and completion of three existing paradigms. Prediction markets offer mature information aggregation and incentives but apply only to short-term, resolvable events; digital twins offer mature large-scale simulation and data assimilation but handle only physical systems without social subjects; global-brain and collective-consciousness research offers the conceptual vision of planetary intelligence \citep{heylighlen2011} but has long lacked an operational engineering path. The initiative adds the temporal dimension to prediction markets through long-horizon automated resolution, adds the social dimension to digital twins through agent-based social simulation, and adds the dimension of power to the global brain through reflexivity governance. Only when these three loops close does the "planetary collective prediction system" become, for the first time, a discussable engineering object. The comparisons of different projects are shown in Table~\ref{tab:save2050-3}.

\begin{table}[H]
\centering
\small
\caption{"Save 2050" compared with existing paradigms.}
\label{tab:save2050-3}
\begin{tabularx}{\linewidth}{YYYYYY}
\toprule
 Paradigm & Aggregation \& incentives & Large-scale simulation & Social subjects & Long-horizon resolution & Governance       \\
\midrule
 Prediction markets & \checkmark{} Mature & \(\times\) & Partial (traders) & \(\times\) Short-term only & Weak             \\
 Digital twins & \(\times\) & \checkmark{} Mature & \(\times\) & --- & ---                \\
 Global-brain research & Conceptual & \(\times\) & \checkmark{} & \(\times\) & Conceptual       \\
 Save 2050 & \checkmark{} & \checkmark{} & \checkmark{} & \checkmark{} (to be built) & \checkmark{} (to be built)  \\
\bottomrule
\end{tabularx}
\end{table}

\subsection{A Phased Roadmap}

\begin{figure}[H]
\centering
\includegraphics[width=\linewidth]{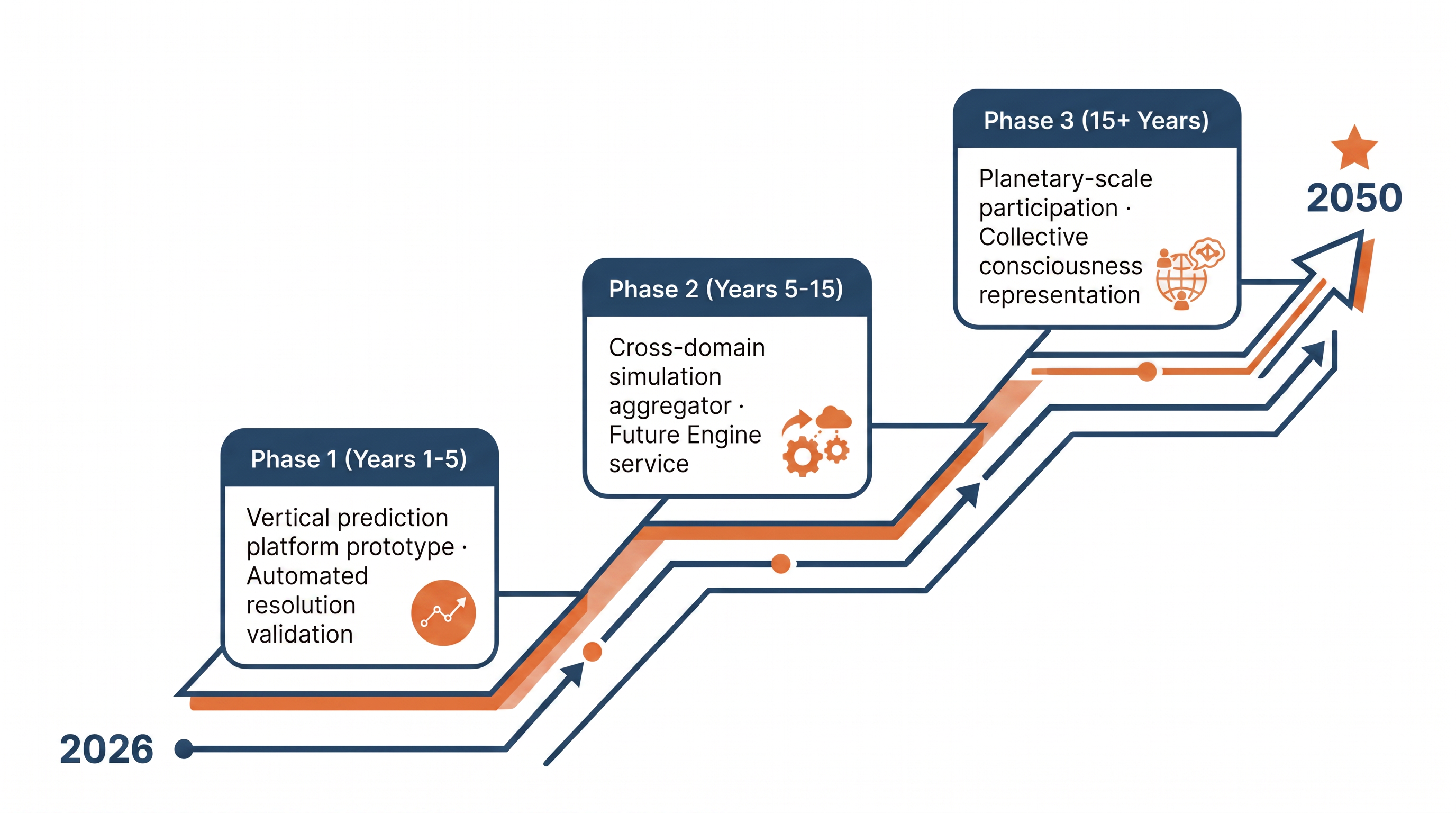}
\caption{The phased roadmap (2026--2050).}
\label{fig:save2050-4}
\end{figure}

We recommend three phases as shown in Figure~\ref{fig:save2050-4}. Phase 1 (near term, 1--5 years): build prediction-platform prototypes in one or two vertical domains (e.g., climate-related events, technology trends), focusing on automated formalization and resolution of natural-language predictions, and validating open crowdsourcing and incentives. Phase 2 (medium term, 5--15 years): build a cross-domain simulation-based aggregator, ingest resolved predictions into the simulation platform, provide "Future Engine" query services, and operate long-horizon prediction contracts within compliance frameworks. Phase 3 (long term, 15+ years): scale participation to the planetary level, complete the reflexivity-governance mechanisms, make the 2050 "predicted world" a continuously evolving object of public scrutiny, and explore institutionalized forms of collective-consciousness representation.

\subsection{Open Problems}

Several theoretical and engineering problems remain open: (1) a rigorous theory of simulation-based aggregation---how to define mathematically the superposition and collapse of predictions as "possibility distributions" in coupled systems, especially the fusion of cross-domain joint distributions; (2) the limits of long-horizon incentives---the boundary of achievable honest incentives under temporal discounting and the absence of anchors; (3) the measurement of collective consciousness---how to detect and measure "emergent patterns of the whole" so that the fidelity of the organ of representation can be tested; (4) verifiable governance---how to make "whether representation is faithful" a publicly checkable question without introducing a centralized authority.

\subsection{Conclusion}

The essence of Save 2050 is to install a "forward-looking system" for human civilization. A civilization speeding through a dark forest needs not greater speed but the ability to see the road ahead. The initiative's primary goal is prediction---faithfully aggregating humanity's judgments about the future. Yet it clearly foresees that when predictions become accurate enough and trusted enough, they may transcend description and become prophecy. Whether the initiative ultimately "saves" 2050 or "kidnaps" it therefore depends on whether the organ that presents the future belongs to everyone---or only to its builders. To make the future visible, discussable, and co-writable: this is a technical problem, a civilizational problem, and a compulsory question for the next two decades.

\bibliographystyle{plainnat}
\bibliography{reference}

\end{document}